\documentclass[twoside]{IEEEtran}
\usepackage[margin=0.75in]{geometry}               
\usepackage{graphicx}
\usepackage{amssymb}
\usepackage{epstopdf}
\title  {Posterior Matching Scheme for Gaussian \\
Multiple Access Channel with Feedback } 
\author{\IEEEauthorblockN{Lan V. Truong},
\emph{Member, IEEE,}\\
\IEEEauthorblockA{Information Technology Specialization Department (ITS),\\
FPT University, Hanoi, Vietnam\\
E-mail: lantv@fpt.edu.vn}}
\date{}                                      
\begin{document}
\maketitle {}
\begin{abstract}                                                           
Posterior matching is a method proposed by Ofer Shayevitz and Meir Feder to design capacity achieving coding schemes for general point-to-point memoryless channels with feedback. In this paper, we present a way to extend posterior matching based encoding and variable rate decoding ideas for Gaussian MAC with feedback, referred to as \emph{time-varying posterior matching scheme}, analyze the achievable rate region and error probabilities of the extended encoding-decoding scheme. The \emph{time-varying posterior matching scheme} is a generalization of the Shayevitz and Feder's posterior matching scheme when the posterior distributions of the input messages given output are not fixed over transmission time slots. It turns out that the well-known Ozarow's  encoding scheme, which obtains the capacity of two-user Gaussian channel, is a special case of our extended posterior matching framework as the Schalkwijk-Kailath's scheme is a special case of the point-to-point posterior matching mentioned above. Furthermore, our designed posterior matching also obtains the linear-feedback sum-capacity for the symmetric multiuser Gaussian MAC.  Besides, the encoding scheme in this paper is designed for the real Gaussian MAC to obtain that performance, which is different from previous approaches where encoding schemes are designed for the complex Gaussian MAC.  More importantly, this paper shows potential of posterior matching in designing optimal coding schemes for multiuser channels with feedback.
\end{abstract}   
\section {Introduction}                        

In his early work [9], Shannon proved that feedback could not increase the capacity of a point-to-point memoryless channel. However, feedback could improve error performance and simplify the transmission scheme for this kind of channel. In [10], Horstein proposed a simple sequential transmission scheme, which achieves the capacity of Binary Symmetric Channel (BSC) and provides larger error exponents than traditional fixed length block coding. Besides, Schalkwijk and Kailath also showed that feedback could improve error performance and/or simplify the transmission scheme for the point-to-point Gaussian channel [7], [8]. For Gaussian multiuser channels, the situation is more interesting. In [12], Gaarder and Wolf proved that feedback can enlarge the capacity region of the multiple access channel, and Ozarow [3] successfully constructed a simple coding scheme for the two user Gaussian MAC with feedback and reaffirmed that feedback could increase the capacity of the channel. Furthermore, Kramer devised a code for complex Gaussian channel based on a beautiful property of the circulant matrix that has all columns of the DFT (Discrete Fourier Transform) matrix as its eigenvectors [14]. This code was proved to obtain the linear-feedback sum-capacity of the symmetric Gaussian channel with feedback in [16]. By using the control-theoretic approach to communications with feedback, Ardestanizadeh and Fraceschetii [17] also proposed a linear code that has the same performance as Kramer's code for symmetric Gaussian complex channels. 

Recently, Shayevitz and Feder [1], [2], and [4] have discovered an underlying principle between the Horstein and Schalkwijk-Kailth schemes in a simple encoding scheme called posterior matching scheme for general point-to-point memoryless channels. The idea of posterior matching is that the transmitter encapsulates the information the receiver does not know up to present time in one random variable and then transmits that random variable to the receiver in the next transmission to refine the receiver's knowledge. The distribution of that variable will be selected in a way such that the input constraint is satisfied. Later, Bae and Anastasopolous extended this scheme for the finite-state channel with feedback by using another approach [11]. Ma and Coleman provided a viewpoint on posterior matching from stochastic control perspectives [18] and generalized this encoding scheme to higher dimension via optimal transportation [19]. One interesting open problem is to extend the Shayevitz and Feder posterior matching scheme for multiuser cases. In this paper, using the same approach as Shayevitz and Feder used for point-to-point  memoryless channels, we propose a posterior matching based encoding and decoding strategy for real Gaussian MACs, referred to as \emph{a time-varying posterior matching scheme}, and analyze the error probabilities for all encoding-decoding schemes designed by using these strategies. 

We analyze the achievable rate region and error performance of encoding and decoding schemes using these strategies by defining a generalized iterated function systems (GIFS) which has the \emph{generalized average contractive} property (asymptotically average contractive). Refer to our Theorem I for more details. Note that our imposed constraint is less strict than the constraint that Shayevitz and Feder imposed to analyze the point-to-point memoryless channels. Specifically, in Theorem 6 in [4], Shayevitz and Feder used the relations between the information rates and contraction properties of the iterated function system (IFS) to analyze the error probability for point-to-point cases. For the continuous cases, they assumed that the reverse iterated function system (RIFS), generated by the kernel $w_y(\cdot):= F_{X|Y}^{-1}\circ F_X$ and controlled by the identically distributed output sequence $\{Y_k\}_{k=1}^{\infty}$, has the average contractive property to analyze the error performance of their posterior matching schemes. That assumption requires the distribution at the output of the point-to-point memoryless channel be identically distributed when using their proposed encoding schemes. This also means that if the output distribution is not identically distributed, the error analysis in Theorem 6 in [4] cannot apply. For example, this situation happens with our proposed matching schemes for the Gaussian MAC in this paper.

Finally, we illustrate our strategies by designing an encoding scheme that obtains optimal performance for the Gaussian MAC. Specifically, our proposed code obtains the same performance as Ozarow's code [3] for the general two-user Gaussian channel, so it achieves the capacity of this channel. For the case when the number of users is greater than 3, our proposed code obtains the same performance as the Kramer's code in the sense of sum-rate, so it is optimal among linear code with respect to sum rate capacity. To the best of our knowledge, the \emph{time-varying posterior matching} in this paper is the first code designed for the real symmetric Gaussian MAC to achieve the linear-feedback sum-capacity when the number of users is greater than 3.

The rest of this paper is organized as follows. Section II presents the channel model and some mathematical preliminaries. Sections III, IV introduce the time-varying posterior matching idea, and perform the error analysis of an encoding-decoding scheme for the Gaussian MAC with feedback constructed by using that idea. A time-varying encoding-decoding strategy and error analysis for the general two-user white Gaussian MAC and the multiuser symmetric white Gaussian MAC are placed in Section V.  Finally, Section VI concludes this paper. 
\section {Channel Model and Preliminaries} 
\subsection{Mathematical notations}
Upper-case letters, their realizations by corresponding lower-case letters, denote random variables. A real-valued random variable $X$ is associated with a distribution $\mathbb{P}_X(\cdot)$ defined on the usual Borel $\sigma$-algebra over $\mathbb{R}$, and we write $X \sim \mathbb{P}_{X}$. The cumulative distribution function (c.d.f.) of $X$ is given by $F_X(x)=\mathbb{P}_X((-\infty,x])$, and their inverse c.d.f is defined to be $F_X^{-1}(t):= \mbox{inf}\{x:F_X(x) > t\}$. The uniform probability distribution over $(0,1)$ is denoted through $\mathcal{U}$. The composition function $(f\circ g)(x)=f(g(x))$. In this paper, we use the following lemma:

{\bf Lemma I:} Let $X$ be a continuous random variable with $X\sim \mathbb{P}_X$ and $\Theta$ be an uniform distribution random variable, i.e. $\Theta \sim \mathcal{U}$ be statistical independent. Then $F_X^{-1}(\Theta) \sim \mathbb{P}_X$ and $F_X(X) \sim \mathcal{U}$.  
\begin{IEEEproof}
Refer to [4] for the proof.
\end{IEEEproof}

Big O notation (with a capital letter O, not a zero), also called Landau's symbol, is a symbolism used in complexity theory, computer science, and mathematics to describe the asymptotic behavior of functions. Basically, it tells you how fast a function grows or declines. For the formal definition, suppose $f(n)$ and $g(n)$ are two functions defined on positive integer number. We write
\[
f(n)=O(g(n))
\] (or $f(n)=O(g(n))$ for $n \rightarrow \infty$ to be more precise) if and only if there exists constants $N$ and $C$ such that
\[
|f(n)| \leq C|g(n)| \hspace{3mm} \mbox{for all} \hspace{1mm} n >N
\] 
Intuitively, this means that $f$ does not grow faster than $g$. 

In addition to big O notations, another Landau symbol is used in mathematics: the little o. Formally, we write $f(n)=o(g(n))$ for $n \rightarrow \infty$ if and only if for every $C>0$ there exists a real number $N$ such that for all $n>N$ we have $|f(n)| < C |g(n)|$. If $g(n) \neq 0$, this is equivalent to $\lim_{n\rightarrow \infty} f(n)/g(n)=0$. 

A Hadamard matrix [15] of order $n$ is an $(n\times n)$ matrix of $+1$s and $-1$s such that ${\bf H}{\bf H}^T=n{\bf I}$. In fact, it is not yet known for which values of $n$ an ${\bf H}_n$ does exists. However, we know that if a Hadamard matrix of order $n$ exists, then $n$ is $1, 2, 4$, or a multiple of $4$. Moreover, if $n$ is of the form $2^m$, $m$ a positive integer,  we can construct ${\bf H}_n$ by using the Sylvester method. Besides,  the Paley construction, which uses quadratic residues, can be used to construct Hadamard matrices of order $n$, where $n$ is of the form $(p+1)$, $p$ is  a prime, and $n$ is a multiple of $4$. 
\subsection{Gaussian Multiple Access Channel with Feedback}
\begin{figure}[!h]
\begin{center}
\includegraphics [scale=0.35] {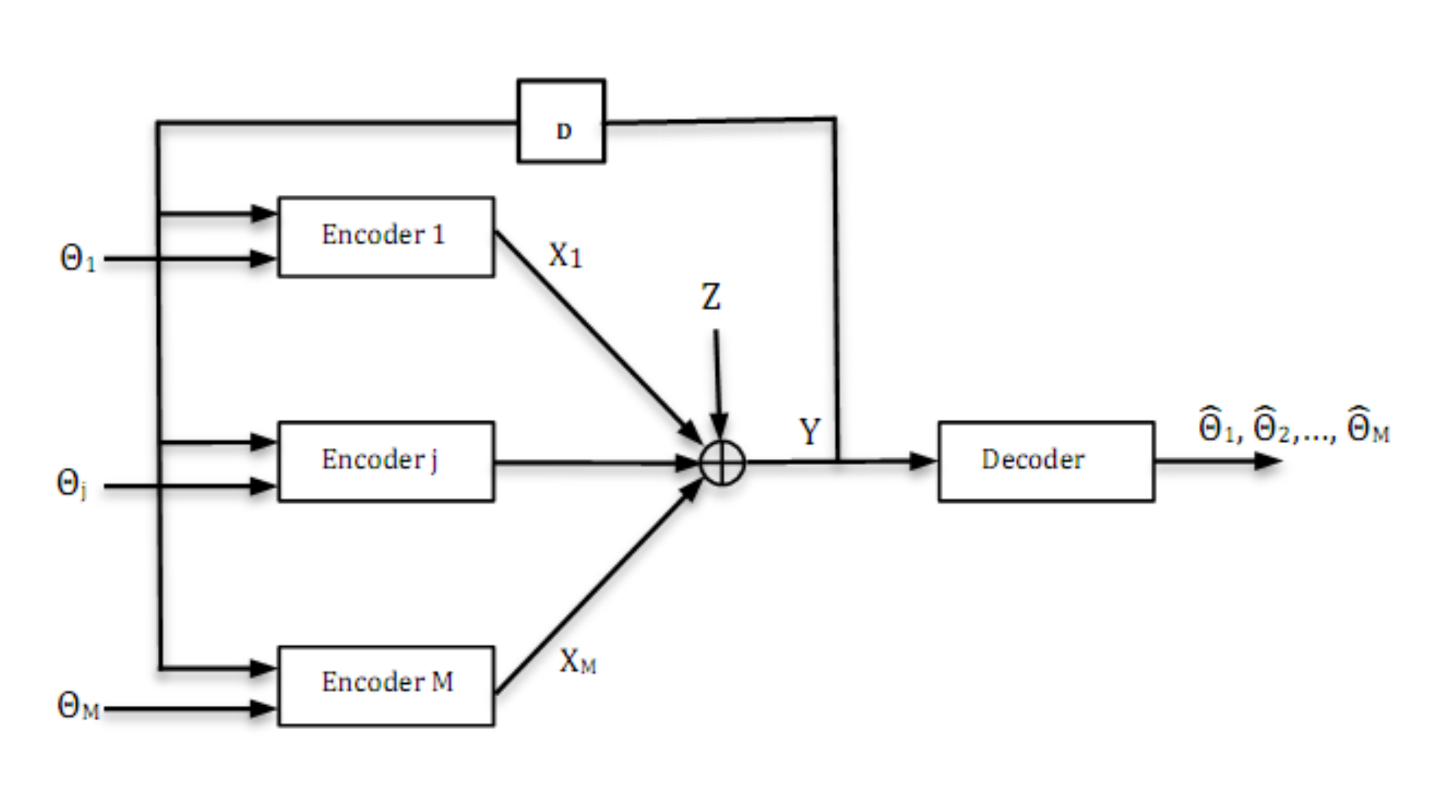}  
\caption{M-user Gaussian MAC with Feedback}
\label{default}
\end{center}
\end{figure}
Consider the communication problem between $M$ senders and a receiver over a multiple access channel with additive Gaussian noise (AWGN-MAC) when channel outputs are noiselessly fed back to all the senders (Figure 1). Each sender $m\in \{1,2,...,M\}$ wishes to reliably transmit a random message point $\Theta_m$, which is uniformly distributed over the unit interval with its binary expansion representing an infinite independent-identically-distributed (i.i.d.) Bernoulli(1/2) sequence, to the receiver. At each time $n$, the output of the channel is
\[
Y_n=\sum_{m=1}^M X_n^{(m)} + Z_n
\]
where $X_n^{(m)} \in \mathbb{R}$ is the transmitted symbol by sender $m$ at time $n$, $Y_n \in \mathbb{R}$ is the output of the channel, and $\{Z_n\}$ is a discrete-time zero mean white Gaussian noise process with unit average power, i.e., $E[Z_n^2]=1$ and is independent of $\Theta_1, \Theta_2,...,\Theta_M$. We assume that output symbols are casually fed back to the sender and the transmitted symbol $X_n^{(m)}$ for sender $m$ at time $n$ can depend on both the message $\Theta_m$ and the previous channel output sequence $Y^{n-1}: =\{Y_1,Y_2,...,Y_{n-1}\}$.

A \emph{transmission scheme} for a Gaussian MAC is a set of $M$ sequences of transmission functions $g^{(m)}_n:(0,1)\times \mathbb{R}^{n-1}\rightarrow \mathbb{R}$ for $m=1,2,\cdots, M$, so that the input to the channel generated by the transmitter is given by
\[
X^{(m)}_n=g^{(m)}_n\left(\Theta_m,Y^{n-1}\right)
\] 

A \emph{decoding rule} for a MAC is set of sequences of measurable mappings $\{\triangle_n^{(m)}: \mathbb{R}^n \rightarrow \mathcal{E} \}_{n=1}^{\infty}$, where $\mathcal{E}$ is the set of all open intervals in $(0,1)$ and $m=1,2,\cdots, M$. Here, $\triangle^{(m)}_n(y^n)$, refers as to the decoded interval for the user $m$. The error probabilities at time $n$ associated with a transmission scheme and a decoding rule, is defined as
\[
p^{(m)}_n(e) := \mathbb{P}\left(\Theta_m \notin \triangle^{(m)}_n(Y^n)\right), \forall m \in \{1,2,...,M\}
\]
and the corresponding achievable rate vector at time $n$ is defined to be
\[
\left\{\left(R^{(1)}_n,R^{(2)}_n,...,R^{(M)}_n\right): R^{(m)}_n =-\frac{1}{n}\log\left|\triangle^{(m)}_n(Y^n)\right|\right\}
\]

We say that a transmission scheme together with a decoding rule achieve a rate vector $(R_1,R_2,...,R_M)$ over a Gaussian MAC if for all $m\in \{1,2,...,M\}$ we have
\begin{equation}
\lim_{n\rightarrow \infty}\mathbb{P}\left(R^{(m)}_n<R_m\right)=0, \lim_{n \rightarrow \infty}p_n^{(m)}(e)=0
\end{equation}

The rate vector is achieved within input power constraints $\left\{P_m\right\}_{m=1}^M$, if in addition
\begin{equation}
\lim_{n\rightarrow \infty}\frac{1}{n}\sum_{k=1}^n E(X^{(m)}_k)^2 \leq P_m, \hspace{2mm} \forall m
\end{equation}

An optimal fixed rate decoding rule for a MAC with rate region $\left(R_1,R_2,...,R_M\right)$ is one that decodes a vector of fixed length intervals $\{(J_1,J_2,...,J_M): |J_m|=2^{-nR_m}, \forall m\}$, whose marginal posteriori probabilities are maximal, i.e.,
\[
\triangle^{(m)}_n(y^n)=\mbox{argmax}_{J_m \in \mathcal{E}: |J_m|=2^{-nR_m}} \mathbb{P}_{\Theta_m|Y^n}(J_m|y^n)
\]

An optimal variable rate decoding rule with target error probabilities $p^{(m)}_e(n)=\delta^{(m)}_n$ is one that decodes a vector of minimal-length intervals $(J_1,J_2,...,J_M)$ with accumulated marginal posteriori probabilities exceeds corresponding targets, i.e.,
\[
\triangle^{(m)}_n(y^n)=\mbox{argmin}_{J_m\in \mathcal{E}: \mathbb{P}_{\Theta_m|Y^n}(J_m|y^n)\geq 1- \delta^{(m)}_n}|J_m|
\]

Both decoding rules make use of the marginal posterior distribution of the message point $\mathbb{P}_{\Theta_m|Y^n}$ which can calculate online at the transmitter $m$ and the receiver.  Refer [4] for more details. A proof that the achievability in the sense of $(1)$ and $(2)$ implies that the achievability in the standard framework are in the Appendix. \\

{\bf Lemma II:} The achievability in the definition $(1)$ and $(2)$ implies the achievability in the standard framework.
\begin{IEEEproof}
Refer to the Appendix.
\end{IEEEproof}

\section{Time-varying Posterior Matching Scheme}
\subsection{Shayevitz and Feder's Posterior Matching Scheme}
In this part, we firstly review the posterior matching scheme proposed by Ofer Shayevitz and Meir Feder for point-to-point channel in [4]. Specifically, the authors argued that after the receiver observed the output sequence $Y^n$, there is still some  "missing information" that can be encapsulated in a random variable $U$ with the following properties: \\

(i) $U$ is statistically independent of $Y^n$.\\

(ii) The message point $\Theta_0$ can be a.s. uniquely recovered from $(U, Y^n).$\\

With that line of thought, they proposed a principle for generating the next channel input as follow: \\

\emph{The transmission function $g_{n+1}$ should be selected so that $X_{n+1}$ is $\mathbb{P}_X$-distributed, and is a fixed function of some random variable $U$ satisfying properties $(i)$ and $(ii)$.}\\

{\bf Lemma III:} (\emph{Posterior Matching Scheme [4]}). The following transmission scheme satisfies the posterior matching principle for any $n$:
\[
g_{n+1}(\theta, y^n)=F_{X}^{-1}\circ F_{(\Theta_0|Y^n)}(\theta |y^n)
\]
Based on the transmission functions defined in section II-B [4],  the input to the channel is a sequence of random variables given by
\begin{equation}
X_{n+1}=F_{X}^{-1}\circ F_{\Theta_0|Y^n}(\Theta_0|Y^n)
\end{equation}
\begin{IEEEproof}
Refer to [4] for the proof.
\end{IEEEproof}
\vspace{2mm}
\subsection{Time-varying Posterior Matching Scheme}

In this section, we propose a posterior matching scheme for additive Gaussian multiple access channel (MAC) with feedback, called \emph{time-varying posterior matching}. Our encoding proposal is based on the  following lemma: \\

{\bf Lemma IV:} For an additive white Gaussian MAC with feedback having $M$ inputs and one output, let the output signal be a linear combination with known coefficients of the input signals, i.e. $Y_n=\sum_{m=1}^M \alpha_n^{(m)} X_n^{(m)}+Z_n$ where $Z_n$ is the additive white Gaussian noise and the coefficients $\alpha_n^{(m)}$ are part of the coding scheme, but are viewed as part of the channel for the purpose of deriving the posterior matching rule. Also assume that the covariance matrix among transmitted symbols at each time slot $n$ defined as
\begin{equation}
{\bf R}_n=\left[ \begin{array}{cccc}E[X_n^{(1)} X_n^{(1)}] &\cdots&E[X_n^{(1)}X_n^{(M)}]\\E[X_n^{(2)}X_n^{(1)}] &\cdots&E[X_n^{(2)}X_n^{(M)}]\\ \vdots&\ddots&\vdots\\E[X_n^{(M)}X_n^{(1)}] &\cdots&E[X_n^{(M)}X_n^{(M)}]\end{array}\right]
\end{equation}
Then the posterior matching scheme in Lemma III at each transmitter $m$, i.e.  
\[
X_{n+1}^{(m)}=F_{X_m}^{-1}\circ F_{(\Theta_{m}|Y^{n})}(\theta_m |y^{n})
\]
becomes a \emph{time-varying posterior matching} given by
\[
X_1^{(m)}=F_{X_m}^{-1}\left(\Theta_m\right); \hspace{2mm} X^{(m)}_{n+1}=F_{X_m}^{-1}\circ F_{X^{(m)}_{n}|Y_n}\left(X^{(m)}_n|Y_n\right)		  
\]
where $X_m$ is  a random variable encapsulating the input power constraint, $X_m \sim \mathcal{N}(0,P_m)$, and $\Theta_m$ is the intended transmitted message at the transmitter $m$. 

In addition, the correlation matrix among transmitted symbols at each time slot $n$, i.e. ${\bf R}_n$, can be calculated online at the transmitters and the receiver. Especially, the posterior distribution $\{X^{(m)}_n|Y_n\}$ can be calculated online at both the transmitter $m$ and the receiver. 
\begin{IEEEproof}
To begin with, we show that $X_n^{(m)}-Y_n-Y^{(n-1)}$ constitutes a Markov chain. Indeed, 
\begin{itemize}
\item From the Lemma I, we know that $F_{\Theta_m|Y^{n-1}}(\Theta_m|y^{n-1}) \sim \mathcal{U}$ and this holds for all $y^n$. Then, 
$ F_{\Theta_m|Y^{n-1}}(\Theta_m|Y^{n-1}) \sim \mathcal{U} $ and is statistically independent of $Y^{n-1}$. Therefore, $X_n^{(m)}$ is independent of $Y^{n-1}$.
\item Since the resulting scheme is linear by our assumption, and the channel is additive Gaussian,  $(X_n^{(1)}, X_n^{(2)},\cdots, X_n^{(m)}, Y^n)$ are jointly Gaussian. 
\end{itemize}
The two observations imply that $(X_n^{(1)}, X_n^{(2)}, \cdots, X_n^{(m)})$ are mutually independent of $Y^{n-1}$, which together with the memoryless property of the channel, implies the Markov chains.

Moreover, by our construction of the multi-letter posterior matching formula, we have
 \[
 g_n^{(m)}(\theta_m,y^{n-1}) = F_{X_m}^{-1}\circ F_{(\Theta_m|Y^{n-1})}(\theta_m |y^{n-1})
 \]
Since $X_m, \Theta_m$ are continuous random variables, their c.d.f. and inverse c.d.f. functions are continuous, hence the composite function $g_n^{(m)}(\theta_m,y^{n-1})$ is continuous on $\theta_m$. Besides, the monotonicity of this function is originated from the monotonicity of the c.d.f. and the inverse c.d.f. 

Now,  let's return to prove the Lemma IV.  Note that
\[
F_{\Theta_m|Y^n}(\theta_m|y^n)=\mathbb{P}(\Theta_m\leq \theta_m| Y^n=y^n)
\]
\[
=\mathbb{P}(g^{(m)}_n(\Theta_{m},y^{n-1})\stackrel{(a)}{\leq} g^{(m)}_n(\theta_m, y^{n-1})|Y^n=y^n)
\]
\[
=\mathbb{P}(X^{(m)}_n \leq g^{(m)}_n(\theta_m, y^{n-1})|Y^n=y^n)
\]
\[
\stackrel{(b)}{=}\mathbb{P}(X^{(m)}_n \leq g^{(m)}_n(\theta_m, y^{n-1})|Y_n=y_n)
\]
\[
=F_{X^{(m)}_n|Y_n}(g^{(m)}_n(\theta_m,y^{n-1})|y_n)
\]
where (a) follows from the fact that the transmission function is continuous and monotone, and (b) follows from the aforementioned fact that $X_n^{(m)}  - Y_n - Y^{(n-1)}$ forms a Markov chain for any $m$. 

Finally, we have
\[
X^{(m)}_{n+1}=F_{X_m}^{-1}\circ F_{\Theta_m|Y^n}(\Theta_m|Y^n)
\]
\[
=F_{X_m}^{-1}\circ F_{X^{(m)}_n|Y_n}(g^{(m)}_n(\Theta_m,Y^{n-1})|Y_n)
\]
\[
=F_{X_m}^{-1}\circ F_{X^{(m)}_n|Y_n}(X^{(m)}_n|Y_n)
\]

Moreover, with \emph{time-varying posterior matching} transmission applied for Gaussian MAC, $X_{n+1}^{(m)}$ is a linear combination of $X_n^{(m)}, Y_n$ (as we will see in the proof of the Theorem II below), so $X_{n+1}^{(m)}$ is a linear combination of $X_n^{(m)}, Z_n, \hspace{1mm} m\in \{1,2,...M\}$. Therefore, the correlation between $X_{n+1}^{(m)}$ and $X_{n+1}^{(k)}$ only depends on the correlations among transmitted symbols at time $n$, i.e. $X_n^{(m)}, \hspace{1mm} m\in \{1,2,...,M\}$. In other words, ${\bf R}_{n+1}$ is a function of ${\bf R}_n$, so the transmitters and receiver can calculate the matrix ${\bf R}_n$ online at both transmitters and receiver. Moreover, from the relation $Y_n=\sum_{m=1}^M \alpha_n^{(m)} X_n^{(m)}+Z_n$, we see that the distribution of $X_n^{(m)}|Y_n$ is a function of all elements in the correlation matrix ${\bf R_n}$. That concludes the proof. 
\end{IEEEproof}
\emph{Remark:} We refer the transmission scheme in Lemma IV to as \emph{time-varying posterior matching scheme}. For a continuous point-to-point memoryless channel, the distribution $X_n|Y_n$ doesn't depend on $n$, hence we have the posterior matching scheme for this case like the formula (16) in [4].  However, in a MAC  (for example additive white Gaussian MAC), where each received signal is a linear combination of all transmitted signals and Gaussian noise, the distribution $X^{(m)}_n|Y_n$ between the input $m$ and the output may be dependent on $n$. Therefore, \emph{time-varying posterior matching scheme} may be the solution to overcome this problem. However, we will see from our proof in the Theorem I below the variable rate decoding rule, or \emph{Generalized Reverse Iterated Function System (GRIFS)}, can be applied at receiver to decode signals if and only if all the distributions $X^{(m)}_n|Y_n$ can be calculated online at the corresponding transmitters and receiver. With the result in the Lemma IV, this condition is always satisfied when \emph{time-varying posterior matching encoding schemes} used at transmitters.  We will show in the next parts that using \emph{time-varying posterior matching scheme} at transmitters can obtain optimal performances for some known cases. 

\section{ Error analysis for time-varying posterior matching scheme}
In this section, we analyze error performance for Gaussian MAC with feedback employing the \emph{time-varying posterior matching scheme} at the transmitters and variable-decoding rule at the receiver.\\

{\bf Theorem I:}  Consider a real Gaussian MAC with $M$ transmitters and one receiver without input power constraints. Assuming that at each transmitter $m$, transmitted sequence $X_n^{(m)}$ conforms to the \emph{time-varying posterior matching rule}, as following:
	 \[
	  X_1^{(m)}=F_{X_m}^{-1}\left(\Theta_m\right); X^{(m)}_{n+1}=F_{X_m}^{-1}\circ F_{X^{(m)}_{n}|Y_n}\left(X^{(m)}_n|Y_n\right)
	 \]
where $X_m \sim N(0,P_m)$ is a Gaussian random variable,  and $Y_n$ is the output of the Gaussian MAC which is a linear combination of all these transmitted signals at time $n$. Let $ w^{(m)}_{Y_n} :=F^{-1}_{X^{(m)}_n|Y_n}\left(\cdot|Y_n\right)\circ F_{X_m}, \hspace{2mm} \forall n\in \mathbb{N}$ and $\displaystyle{L_{s,t}(h):=\frac{|h(s)-h(t)|}{|s-t|}} $ as the global Lipschitz operator.  Under the conditions that
\begin{equation}
0< \limsup_{n\rightarrow \infty}\left(\mbox{sup}_{\tiny {s\neq t}} E\left[L_{s,t}\left(w^{(m)}_{Y_n}\right)\right]\right)  <1
\end{equation}
define: 
$
R_m^{*}:=-\limsup_{n\rightarrow \infty }\log\left(\mbox{sup}_{\tiny {s\neq t}} E\left[L_{s,t}\left(w^{(m)}_{Y_n}\right)\right]\right)
$
Then the rate region $\{(R_1,R_2,...,R_M): R_m < R_m^{*}\}$ is achievable and the error probabilities $p^{(m)}_n(e)$ decay to zero as 
	  \[
	   -\log p^{(m)}_n(e) =\circ\left(2^{2n(R^*_m-R_m - \delta)}\right), \hspace{2mm} \forall m \in \{1,2,...,M\}
           \] 
\begin{IEEEproof}
First, observe that since the distributions $X_n^{(m)}|Y_n, n=1,2,...$ can be calculated online at both transmitters and receiver by Lemma IV, therefore $w^{(m)}_{Y_n}$ can be calculated online at both the transmitters and receiver.  Denote $T^{(m)}_n(s):= w^{(m)}_{Y_1}\circ w^{(m)}_{Y_2}... \circ w^{(m)}_{Y_n}(s), \hspace{2mm} \forall s\in \mathbb{R}$,  referred as to a \emph{Generalized Iterated Function System (GIFS)} generated by the \emph{kernel} sequence $\{w^{(m)}_{Y_n}\}_{n=1}^{\infty}$. For each $m \in \{1,2,...,M\}$, select a fixed interval $J_1^{(m)}=(s_m,t_m) \subset \mathbb{R}$ as the decoded interval with respect to $X_{n+1}^{(m)}$.  

Define the corresponding interval at the origin to be $J_n^{(m)}:=\left(T^{(m)}_n(s_m),T^{(m)}_n(t_m)\right)$ and set them to be the decoded interval for $X_1^{(m)}=F_{X_m}^{-1}\left(\Theta_m\right)$, and so the decoded interval for $\Theta_m$ are set to be $\Delta_n^{(m)}\left(Y^n\right)=F_{X_m}(J_n^{(m)})$. 

It is easy to see that 
\[
R_m^* = \log r_m^{-1}
\]
where
\begin{equation}
r_{m}:=\limsup_{n\rightarrow \infty} r_{m}^{(n)}
\end{equation}
For any fixed rate $R_m < R_m^*$, we can find an $\epsilon >0$ such that $R_m < \log (r_m +\epsilon)^{-1} < R_m^*$. 
Observe that: 
\[
\mathbb{P}\left(R_n^{(m)}< R_m \right)\stackrel{(a)}{=} \mathbb{P}\left( -\frac{1}{n} \log\left|\Delta_n^{(m)}(Y^n)\right| <R_m\right)
\]
\[
= \mathbb{P}\left(|\Delta_n^{(m)}| > 2^{-nR_m}\right)\stackrel{(b)}{=}\mathbb{P}\left[F_{X_m}(J_n^{(m)})> 2^{-nR_m}\right]
\]
\begin{equation}
\leq  \mathbb{P}\left(|J_n^{(m)}|> 2^{-nR_m}/K_m\right)
\end{equation}
where $K_m = \sup_{x \in \mathbb{R}}\{f_{X_m}(x)\}$. Here, (a) follows from the definition of the instant corresponding achievable rate vector at time $n$ in the Section II above, and (b) follows from the fact that we set the decoded interval  $\Delta_n^{(m)}(Y^n)= F_{X_m}(J_n^{(m)})$. \\

On the other hand, since $r_{m}=\limsup_{n\rightarrow \infty} r_{m}^{(n)}<1 $, there exists an  $N_{\epsilon}\in \mathbb{N}$ such that 
$\sup_{n\geq N_{\epsilon}} r_{m}^{(n)} < r_{m}+\epsilon$. Let $v_{m}:=\sup_{1\leq k \leq N_{\epsilon}}\sup_{s\neq t} L_{s,t}(w_{Y_k}^{(m)})$. From (7) we have
\[
\mathbb{P}\left(R_n^{(m)}< R_m \right)\leq \mathbb{P}\left(|J_n^{(m)}|> 2^{-nR_m}/K_m\right)
\]
\[
\stackrel{(a)}{\leq} K_m2^{nR_m}E\left[E\left(|w^{(m)}_{Y_1}\circ w^{(m)}_{Y_2}... \circ w^{(m)}_{Y_n}(s_m)\right. \right.
\]
\[
\left. \left. -w^{(m)}_{Y_1}\circ w^{(m)}_{Y_2}... \circ w^{(m)}_{Y_n}(t_m)||Y_2^n\right)\right]
\]
\[
\stackrel{(b)}{\leq}K_m2^{nR_m}t_{m} E\left(|w^{(m)}_{Y_2}... \circ w^{(m)}_{Y_n}(s_m)\right. 
\]
\[
\left. -w^{(m)}_{Y_2}... \circ w^{(m)}_{Y_n}(t_m)|\right)
\]
\[
\stackrel{(c)}{\leq}K_m2^{nR_m}(v_{m})^{N_{\epsilon}}E\left(|w^{(m)}_{Y_{N_{\epsilon}}}... \circ w^{(m)}_{Y_{n}}(s_m)\right.
\]
\[
\left. -w^{(m)}_{Y_{N_{\epsilon}}}... \circ w^{(m)}_{Y_n}(t_m)|\right)
\]
\begin{equation}
\cdots\stackrel{(d)}{<} K_m2^{nR_m}(v_{m})^{N_{\epsilon}} (r_{m}+\epsilon)^{(n-N_{\epsilon})} |J_1^{(m)}|
\end{equation}
where $(a)$ follows from the Markov's inequality, $(b)$ follows from $v_{m}:=\sup_{1\leq k \leq N_{\epsilon}}\sup_{s\neq t} L_{s,t}(w_{Y_k}^{(m)})$, and $(c)$ is a recursive application of the preceding transitions, $(d)$ follows from $\sup_{n\geq N_\epsilon} r_{m}^{(n)} < r_{m}+\epsilon$ above and recursive applications of the preceding transitions. 

From (8), it is easy to see that a sufficient condition for both $P(R_n^{(m)} < R_m) \rightarrow 0$ is given by choosing  $|J_1^{(m)}|=o\left(2^{n(\log (r_{m}+\epsilon)^{-1}-R_m)}\right)$. Since $\mathbb{P}\left(R_n^{(m)}< R_m \right)\rightarrow 0$ depends only on the length of $J^{(m)}_1$,  so we can choose $s_m=-t_m$. Furthermore, from the Lemma I, we are easy to come to conclusion that $X^{(n)}_m \sim \mathcal{N}(0,P_m)$.  Denote $Q(x)$ as the well-known tail function of the standard normal distribution and use the Chernoff bound of this function we obtain
\[
p^{(m)}_n(e)=\mathbb{P}\left(\Theta_m \notin F_{X_m}\left(J_n^{(m)}\right)\right)
=\mathbb{P}\left(X_{n+1}^{(m)} \notin J_1^{(m)}\right)
\]
\[
=1-\mathbb{P}_{X_m}(J_1^{(m)})\stackrel{(a)}{=} 2 Q\left(\frac{|J_1^{(m)}|}{2} \right)
\]
\begin{equation}
\stackrel{(b)}{=}O\left( \exp\left(-\frac{|J_1^{(m)}|^2}{8 P_m} \right) \right) 
\end{equation}
Here, (a) follows from the fact that $J_1^{(m)}$ is symmetric $s_m = -t_m$, and (b) follows from the Chernoff bound for the Q-function $0 < Q(x) \leq (1/2) \exp(-x^2/2), \hspace{2mm} \forall x > 0$.

To put it simply, any rate vector $\{\left(R_1,R_2,...,R_M\right):R_m < R_m^*,  \hspace{2mm} \forall m \in \{1,2,...,M\}\} $ is achievable, where
\[
 R_m^*= \log r_m^{-1}=-\limsup_{n\rightarrow \infty}\log\left(\sup_{\tiny {s\neq t}}E\left[L_{s,t}\left(w^{(m)}_{Y_n}\right)\right]\right)
\]
The error probabilities decay to zero as 
\[
 -\log p^{(m)}_n(e) =\circ\left(2^{2n(\log(r_m+\epsilon)^{-1}-R_m)}\right)\approx \circ\left(2^{2n(R^*_m-R_m)}\right)
\]

\emph{Remark:} Since we can estimate $R_m^*$ and know our desired rate $R_m$ in advance, it is possible to choose $\epsilon$ by target. This means that the decoding algorithm is technically realizable.  However, there is a tradeoff between the transmission rate $R_m$ (the possible values of $\epsilon$) and the code length $n$. If we transmit at the rate $R_m$ very close to $R_m^*$, we need to choose $\epsilon$ to be very small. As a result, the required $N_{\epsilon}$ may be very big. Furthermore, the fact that $\log (r_m +\epsilon)^{-1}$ is very lose to $R_m$ also makes the error probabilities slowly decayed to zero. To put it simply, the code length $n$ may be very large if we transmit at the rate $R_m$ is nearly $R_m^*$. On the contrary, being able to choose quite large $\epsilon$ makes the required $N_{\epsilon}$ smaller and the decay of error probabilities faster.  
\end{IEEEproof}
\section{A Posterior Matching Scheme for Gaussian MAC with Feedback}
In this section, we consider a real Gaussian MAC with $M$ receivers and input power constraints $P_1,P_2,...,P_M$ at the transmitters $1,2,...,M$, respectively as defined in the section II.  Our encoding scheme for this channel as following: 
\subsection{Encoding}
 \begin{itemize}
\item  At the time interval $n$, each transmitter $m \in \{1,2,...,M\}$ creates a random variable $X_n^{(m)}$ following the posterior matching rule:
\[
X_1^{(m)}=F_{X_m}^{-1}(\Theta_m)
\]
\[
X_{n+1}^{(m)}=F_{X_m}^{-1}\circ F_{X_n^{(m)}|Y_n}\left(X_n^{(m)}|Y_n\right)
\]
where $X_m \sim N(0,P_m)$ is a Gaussian random variable, and $Y_n$ is the output of the MAC. 
\item After that, the transmitter $m$ sends the signal $\alpha_n^{(m)}X_n^{(m)}, $
where ${\bf \alpha}_n=[\begin{array}{cccc}\alpha_n^{(1)}&\alpha_n^{(2)},&\cdots,&\alpha_n^{(M)}\end{array}]^T$
is the column ($n$ mod $M+1$) of the Hadamard matrix $M$ by $M$.
\end{itemize}
\subsection{Decoding}
 \begin{itemize}
 \item At each time slot $n$, the receiver selects a fixed interval $J_1^{(m)}=(s_m,t_m) \subset \mathbb{R}$ as the decoded interval with respect to $X_{n+1}^{(m)}$.
\item Then, set the decoded interval  $J_n^{(m)}=\left(T^{(m)}_n(s_m),T^{(m)}_n(t_m)\right)$  as the decoded interval with respect to $X_1^{(m)}$, where  
 \[
 T^{(m)}_n(s):= w^{(m)}_{Y_1}\circ w^{(m)}_{Y_2}... \circ w^{(m)}_{Y_n}(s), \hspace{2mm} \forall s\in \mathbb{R}
 \]
and $ w^{(m)}_{Y_n} :=F^{-1}_{X^{(m)}_n|Y_n}\left(\cdot|Y_n\right)\circ F_{X_m}, \hspace{2mm} \forall n\in \mathbb{N}$.
\item The receiver sets the decoded interval for the message $\Theta_m$ is
\[
\Delta_n^{(m)}\left(Y^n\right)=F_{X_m}(J_n^{(m)})
\] 
\end{itemize}

We refer this encoding strategy as \emph{Gaussian MAC posterior matching feedback coding and decoding strategy}. \\

{\bf Theorem II}:   Using the  \emph{Gaussian MAC posterior matching feedback coding and decoding strategy} above, the rate region $\{(R_1,R_2,...,R_M): R_m < R_m^{*}\}$ is achievable for Gaussian MAC with feedback, where
\[
R_m^{*}=-\frac{1}{2}\limsup_{n\rightarrow \infty} \left( \log L_n^{(m)}\right)
\]
by setting the target error probabilities
\[
-\log p_n^{(m)}(e)= o\left(2^{2n(R^{*}_m-R_m)}\right), \hspace{2mm} \forall m
\]
where 
\[
L_n^{(m)}= 1- \frac{\left( \sum_{t=1}^M \alpha_n^{(t)}\rho_{n}^{(t,m)}\sqrt{P_t}\right)^2}{\sum_{t=1}^M\sum_{l=1}^M \alpha_n^{(t)}\alpha_n^{(l)}\rho_n^{(t,l)}\sqrt{P_tP_m}+1}
\]
with
\[
\rho_n^{(t,l)} : =\left\{\begin{array}{l}\frac{E[X_n^{(t)}X_n^{(l)}]}{\sqrt{P_t P_l}},\hspace{8mm} t \neq l \\ 1, \hspace{10mm} t=l \end{array}\right.
\]
\begin{IEEEproof}  Applying the Lemma I, we see that  for any $m \in \{1,2,...,M\}$  then:
\begin{equation}
X_n^{(m)} \sim  \mathcal{N}(0,P_m), \hspace{2mm} \forall n \in \mathbb{N}
\end{equation}
Observe that, by this transmission strategy, each transmitter $m,\hspace{1mm} m \in \{1,2,...,M\}$ transmits $\alpha_n^{(m)}X_n^{(m)}$ at time $n$ with $|\alpha_n^{(m)}|=1$,
thus the input power constraints at all transmitters are always satisfied at each transmission time $n,\hspace{1mm} n=1,2...$
Moreover, the output at receiver at time $n$ will be
\[
Y_n= \alpha_n^{(1)} X_n^{(1)}+\alpha_n^{(2)} X_n^{(2)}+....+\alpha_n^{(M)} X_n^{(M)} + Z_n
\]
Thus 
\begin{equation}
\mbox{cov}(X_n^{(m)},Y_n)=\sum_{t=1}^M \alpha_n^{(t)}\rho_n^{(t,m)}\sqrt{P_tP_m}
\end{equation}
\begin{equation}
\mbox{var}(Y_n)=\sum_{t=1}^M\sum_{l=1}^M \alpha_n^{(t)}\alpha_n^{(l)}\sqrt{P_tP_l}\rho_n^{(t,l)}+1
\end{equation}
Then, we have
\[
E[X_n^{(m)}|Y_n]= \frac{\mbox{cov}(X_n^{(m)},Y_n)}{\mbox{var}(Y_n)}Y_n
\]
\begin{equation}
= \frac{\sqrt{P_m}\left(\sum_{t=1}^M \alpha_n^{(t)}\rho_n^{(t,m)}\sqrt{P_t}\right)}{\sum_{t=1}^M\sum_{l=1}^M \alpha_n^{(t)}\alpha_n^{(l)}\sqrt{P_tP_l}\rho_n^{(t,l)}+1}Y_n:=A_n^{(m)}Y_n
\end{equation}
and
\[
\mbox{var}\left(X_n^{(m)}|Y_n\right)=\mbox{var}(X_n^{(m)})-\frac{[\mbox{cov}(X_n^{(m)},Y_n)]^2}{\mbox{var}(Y_n)}
\]
\[
=P_m-\frac{P_m\left(\sum_{t=1}^M \alpha_n^{(t)}\rho_n^{(t,m)}\sqrt{P_t}\right)^2}{\sum_{t=1}^M\sum_{l=1}^M \alpha_n^{(t)}\alpha_n^{(l)}\sqrt{P_tP_l}\rho_n^{(t,l)}+1}:=B_n^{(m)} 
\]
where
\begin{equation}
A_n^{(m)}=\frac{\sqrt{P_m}\left(\sum_{t=1}^M \alpha_n^{(t)}\rho_n^{(t,m)}\sqrt{P_t}\right)}{\sum_{t=1}^M\sum_{l=1}^M \alpha_n^{(t)}\alpha_n^{(l)}\sqrt{P_tP_l}\rho_n^{(t,l)}+1}
\end{equation}
and
\begin{equation}
B_n^{(m)}=P_m-\frac{P_m\left(\sum_{t=1}^M \alpha_n^{(t)}\rho_n^{(t,m)}\sqrt{P_t}\right)^2}{\sum_{t=1}^M\sum_{l=1}^M \alpha_n^{(t)}\alpha_n^{(l)}\sqrt{P_tP_l}\rho_n^{(t,l)}+1} 
\end{equation}
Finally, we obtain
\begin{equation}
F_{X_n^{(m)}|Y_n}\left(X_n^{(m)}|Y_n\right)=\Phi\left(\frac{X_n^{(m)}-A_n^{(m)}Y_n}{\sqrt{B_n^{(m)}}}\right)
\end{equation}
where
\[
\Phi(x)=\frac{1}{\sqrt{2\pi}}\int_{-\infty}^x e^{-t^2/2}dt
\]
Moreover, from $(10)$ we have:
\[
X_n^{(m)} \sim \mathcal{N}(0,P_m)
\]
thus,
\[
F_{X_m}(x)=\Phi\left(\frac{x}{\sqrt{P_m}}\right)
\]
Combining with $(16)$, we obtain:
\[
F_{X_m}^{-1}\circ F_{X_n^{(m)}|Y_n} \left(X_n^{(m)}|Y_n\right)=\sqrt{P_m} \frac{X_n^{(m)}-A_n^{(m)}Y_n}{\sqrt{B_n^{(m)}}} 
\]
Hence,
\[
w_{Y_n}^{(m)}(s)=\sqrt{\frac{B_n^{(m)}}{P_m}}s +A_n^{(m)} Y_n
\]
Finally, we have:
\[
\limsup_{n \rightarrow \infty} \left(\mbox{sup}_{s\neq t}E\left[L_{s,t}\left(w_{Y_n}^{(m)}(s)\right)\right]\right)=\limsup_{n \rightarrow \infty} \sqrt{\frac{B_n^{(m)}} {P_m} }
\]
\[
=\limsup_{n \rightarrow \infty}\left[1-\frac{\left(\sum_{t=1}^M \alpha_n^{(t)}\rho_n^{(t,m)}\sqrt{P_t}\right)^2}{\sum_{t=1}^M\sum_{l=1}^M \alpha_n^{(t)}\alpha_n^{(l)}\rho_n^{(t,l)}\sqrt{P_tP_l}+1}\right]^{1/2}
\]
If we can achieve
\[
0< \liminf_{n \rightarrow \infty} \frac{\left(\sum_{t=1}^M \alpha_n^{(t)}\rho_n^{(t,m)}\sqrt{P_t}\right)^2}{\sum_{t=1}^M\sum_{l=1}^M \alpha_n^{(t)}\alpha_n^{(l)}\rho_n^{(t,l)}\sqrt{P_tP_l}+1}<1
\]
then
\[
0 < \limsup_{n \rightarrow \infty}\left[1-\frac{\left(\sum_{t=1}^M \alpha_n^{(t)}\rho_n^{(t,m)}\sqrt{P_t}\right)^2}{\sum_{t=1}^M\sum_{l=1}^M \alpha_n^{(t)}\alpha_n^{(l)}\rho_n^{(t,l)}\sqrt{P_tP_l}+1}\right] <1
\]

This means that the condition in the Theorem I is satisfied, which leads to  the rate region $\{(R_1,R_2,...,R_M): R_m < R_m^{*}\}$ is achievable, where
\[
R_m^{*}=-\limsup_{n\rightarrow \infty}\log \left(E\left[L_{s,t}\left(w^{(m)}_{Y_n}\right)\right]\right)
\]
\[
=-\limsup_{n \rightarrow \infty} \frac{1}{2}\log\left(\frac{B_n^{(m)}}{P_m}\right)=-\frac{1}{2}\limsup_{n \rightarrow \infty} \left(\log L_n^{(m)}\right)
\]

Note that this capacity region is obtained by setting the target error probabilities $p^{(m)}_n(e) \rightarrow 0$ under the constraints
\[
-\log p_n^{(m)}(e)=o\left(2^{2n(R^{*}_m-R_m)}\right), \hspace{2mm} \forall m \in \{1,2,...,M\}
\]
which have the well-known double-exponential behavior. 
\end{IEEEproof}
\vspace{3mm}
In the following, we will specify achievable rate regions, and error probabilities for two cases: the general two-user Gaussian MAC, and the real symmetric Gaussian MAC when the number of users is arbitrary. \\

From the Theorem II, we see that the strategy to design posterior encoding scheme for multiple access channels with feedback is to find the sequences $\alpha_n^{(m)}$ such that 
\begin{equation}
0< \liminf_{n \rightarrow \infty} \frac{\left(\sum_{t=1}^M \alpha_n^{(t)}\rho_n^{(t,m)}\sqrt{P_t}\right)^2}{\sum_{t=1}^M\sum_{l=1}^M \alpha_n^{(t)}\alpha_n^{(l)}\rho_n^{(t,l)}\sqrt{P_tP_l}+1}<1
\end{equation} for all $m=1,2,...,M$.\\

\emph{Case 1: Two-user Gaussian MAC with feedback}. \\
To show that Ozarow's coding scheme [3] is a special case of our posterior matching framework, we can set $\alpha_n^{(1)}=1, \alpha_n^{(2)}=\mbox{sgn}(\rho_n)$ and later prove that $\mbox{sgn}(\rho_{n+1})=-\mbox{sgn}(\rho_n)$. Observe that the constraint $(17)$ can be also checked to be satisfied by this setting since
\[
0<  \frac{[\sqrt{P_1}\rho_n+\sqrt{P_2}\mbox{sgn}(\rho_n)]^2}{P_1+P_2+2|\rho_n|\sqrt{P_1P_2}+1}<1
\]
for $m=1$ and
\[
 0< \frac{[\sqrt{P_1}+\sqrt{P_2}|\rho_n|]^2}{P_1+P_2+2|\rho_n|\sqrt{P_1P_2}+1}<1
 \]
 for $m=2$. \\

Now we need to find the recursion of $X_n^{(1)}$ and $X_n^{(2)}$ in this case.  Observe that the output sequence:
\begin{equation}
Y_n= X^{(1)}_{n}+X^{(2)}_{n}\mbox{sgn}(\rho_n)+Z_n
\end{equation}
where $Z_n$ is noise process and $Z_n \sim \mathcal{N}(0,1)$. \\
From $(14), (15)$ we have
\[
A_n^{(1)}=\frac{\sqrt{P_1}(\sqrt{P_1}+|\rho_n|\sqrt{P_2})}{P_1+P_2+2|\rho_n|\sqrt{P_1P_2}+1}
\]
\[
A_n^{(2)}=\frac{\sqrt{P_2}(\sqrt{P_2}+\sqrt{P_1}|\rho_n|)\mbox{sgn}(\rho_n)}{P_1+P_2+2|\rho_n|\sqrt{P_1P_2}+1}
\]
and
\[
B_n^{(1)}=P_1\frac{P_2(1-\rho_n^2)+1}{P_1+P_2+2|\rho_n|\sqrt{P_1P_2}+1} 
\]
\[
B_n^{(2)}=P_2\frac{P_1(1-\rho_n^2)+1}{P_1+P_2+2|\rho_n|\sqrt{P_1P_2}+1}
\]
Therefore, we obtain:
\[
X^{(1)}_{n+1}=\sqrt{\frac{P_1+P_2+1+2|\rho_{n}|\sqrt{P_1P_2}}{P_2(1-\rho_{n}^2)+1}}
\]
\[
\times \left(X^{(1)}_{n}-\frac{P_1+|\rho_{n}|\sqrt{P_1P_2}}{P_1+P_2+1+2|\rho_{n}|\sqrt{P_1P_2}}Y_{n}\right)
\]
and
\[
X^{(2)}_{n+1}=\sqrt{\frac{P_1+P_2+1+2|\rho_{n}|\sqrt{P_1P_2}}{P_1(1-\rho_{n}^2)+1}}
\]
\[
 \left(X^{(2)}_{n}-\frac{P_2+|\rho_{n}|\sqrt{P_1P_2}}{P_1+P_2+1+2|\rho_{n}|\sqrt{P_1P_2}}Y_{n}\right)
\] 
\[
\rho_{n+1}:=\frac{E[X_{n+1}^{(1)}X_{n+1}^{(2)}]}{\sqrt{\mbox{var}(X_{n+1}^{(1)})\mbox{var}(X_{n+1}^{(2})}}
\]
\begin{equation}
=\frac{\rho_n-\mbox{sgn}(\rho_n)\sqrt{P_1P_2}(1-\rho_n^2)}{\sqrt{[P_2(1-\rho_n^2)+1][P_1(1-\rho_n^2)+1]}}
\end{equation}\\

Finally, the \emph{time-varying posterior matching encoding} scheme for Gaussian MAC with feedback in this special case as following:\\
\begin{itemize}
\item Step 1:
\[
\rho_1=E[F_{X_1}^{-1}(\Theta_1)F_{X_2}^{-1}(\Theta_2)] =0
\]
Transmitter 1 sends:
\[
X^{(1)}_1=F_{X_1}^{-1}(\Theta_1)
\]
Transmitter 2 sends:
\[
X^{(2)}_1=F_{X_2}^{-1}(\Theta_2)
\]

\item Step $n+1,\hspace{2mm} n \geq 0$,\\
Both transmitters estimate:
\[
\rho_{n+1}=\frac{\rho_n-\mbox{sgn}(\rho_n)\sqrt{P_1P_2}(1-\rho_n^2)}{\sqrt{[P_2(1-\rho_n^2)+1][P_1(1-\rho_n^2)+1]}}
\]
Transmitter 1 sends:
\[
X^{(1)}_{n+1}=\sqrt{\frac{P_1+P_2+1+2|\rho_{n}|\sqrt{P_1P_2}}{P_2(1-\rho_{n}^2)+1}}
\]
\[
\left(X^{(1)}_{n}-\frac{P_1+|\rho_{n}|\sqrt{P_1P_2}}{P_1+P_2+1+2|\rho_{n}|\sqrt{P_1P_2}}Y_{n}\right)
\]
Transmitter 2 sends:
\[
\mbox{sgn}(\rho_{n+1})X^{(2)}_{n+1}=\sqrt{\frac{P_1+P_2+1+2|\rho_{n}|\sqrt{P_1P_2}}{P_1(1-\rho_{n}^2)+1}}
\]
\[
\times \left(X^{(2)}_{n}-\frac{P_2+|\rho_{n}|\sqrt{P_1P_2}}{P_1+P_2+1+2|\rho_{n}|\sqrt{P_1P_2}}Y_{n}\right)\mbox{sgn}(\rho_{n+1})
\] 
\end{itemize}
\vspace{2mm}

{\bf Achievable rate region and error analysis:}\\

For this special case, we have
\begin{equation}
L_n^{(1)}=\frac{P_2(1-\rho_n^2)+1}{P_1+P_2+1+2|\rho_n|\sqrt{P_1P_2}}
\end{equation}
and
\begin{equation}
L_n^{(2)}=\frac{P_1(1-\rho_n^2)+1}{P_1+P_2+1+2|\rho_n|\sqrt{P_1P_2}}
\end{equation}

Apply the Theorem II above, we obtain the achievable rate region for Gaussian MAC with two users as $\left\{(R_1,R_2): R_1< R^{*}_1, R_2< R^{*}_2 \right\}$ 
where 
\begin{equation}
R_1^{*}= \frac{1}{2}\liminf_{n\rightarrow \infty} \log\left(\frac{P_1+P_2+1+2|\rho_n|\sqrt{P_1P_2}}{P_2(1-\rho_n^2)+1}\right)
\end{equation}

Similarly,
\begin{equation}
R_2^{*}=\frac{1}{2}\liminf_{n\rightarrow \infty} \log\left(\frac{P_1+P_2+1+2|\rho_n|\sqrt{P_1P_2}}{P_1(1-\rho_n^2)+1}\right)
\end{equation}
by setting the target error probability to
\[
-\log p_n^{(1)}(e)= o\left(2^{2n(R^{*}_1-R_1)}\right)
\]
and
\[
-\log p_n^{(2)}(e)= o\left(2^{2n(R^{*}_2-R_2)}\right)
\]

Much like Ozarow in $[3]$, at the reception $1$, the receiver adds an independent random variable $W \sim \mathcal{N}(0,\sigma_w^2)$ before feeding back the first receiver signal to the transmitters 1 and 2 to set $|\rho_2|=\rho^*$, where $\rho^*$ is the biggest solution in $(0,1)$ of the following equation:
\begin{equation}
\rho+\frac{\rho -\sqrt{P_1P_2}(1-\rho^2)}{\sqrt{[P_2(1-\rho^2)+1][P_1(1-\rho^2)+1]}}=0
\end{equation}

By this changing, from $(19$ we see that  $\rho_n=(-1)^{n+1}\rho^*$, so $\mbox{sgn}(\rho_{n+1})=-\mbox{sgn}(\rho_n)$ as mentioned above. We also have $\liminf_{n\rightarrow \infty}|\rho_n|=\rho^*$, where $\rho^*$ is a positive solution of the equation $(24)$. 
Replace this result to $(22),(23)$ and combine with $(24)$, we have:
\[
R_1^{*}=\frac{1}{2}\log\left[1+P_1(1-(\rho^*)^2)\right] 
\]
\[
R_2^{*}= \frac{1}{2}\log\left[1+P_2(1-(\rho^*)^2)\right]
\]
\[
R_1^*+R_2^*=\frac{1}{2}\log\left(1+P_1+P_2+2\rho^*\sqrt{P_1P_2}\right)
\]
where $\rho^*$ is defined above. We see that, all the results are the same as Ozarow's results in [3]. So our posterior matching encoding scheme is optimal for Gaussian channel MAC with two users.\\

\emph{Case 2: M-user symmetric Gaussian MAC with feedback}. We consider symmetric case, where $ P_1=P_2=...=P_M=P$.\\

{\bf Achievable rate region and error analysis:} 

Assuming that all the transmitted messages are statistically independent.  Define the normalized covariance matrix by
\[
{\bf R}_n=\frac{1}{P}E\left[{\bf X}_n{\bf X}_n^T\right]
\]
Then
\[
{\bf R}_n=\frac{1}{P}\left[ \begin{array}{cccc}E[X_n^{(1)} X_n^{(1)}] &\cdots&E[X_n^{(1)}X_n^{(M)}]\\E[X_n^{(2)}X_n^{(1)}] &\cdots&E[X_n^{(2)}X_n^{(M)}]\\ \vdots&\ddots&\vdots\\E[X_n^{(M)}X_n^{(1)}] &\cdots&E[X_n^{(M)}X_n^{(M)}]\end{array}\right]
\]
\[
=\left[ \begin{array}{cccc}\rho^{(1,1)}_n&\cdots&\rho^{(1,M)}_n\\\rho^{(2,1)}_n&\cdots&\rho^{(2,M)}_n\\ \vdots&\ddots&\vdots\\\rho^{(M,1)}_n &\cdots&\rho^{(M,M)}_n\end{array}\right]
\]
where  
\[
\rho^{(m,k)}_n:=\frac{E[X_n^{(m)}X_n^{(k)}]}{\sqrt{\mbox{var}(X_n^{(m)})\mbox{var}(X_n^{(k)}})}=\frac{E[X_n^{(m)}X_n^{(k)}]}{P}
\]
is the correlation coefficient between $X_n^{(m)}$ and $X_n^{(k)}$.\\

We will prove by induction that the normalized covariance  has all the columns of the Hadamard matrix $(M$ by $M)$ as its eigenvectors and that ${\bf R}_n$ is symmetric positive definite for all $n=1,2,3,...$

Indeed, with the assumption all the transmitted information messages are statistically independent, we will have ${\bf R}_1={\bf I}_M$, which is an identity matrix of size $M$. Therefore, it is obvious that all the columns of the Hadamard matrix $(M$ by $M)$ are eigenvectors of the matrix ${\bf R}_1$ and that ${\bf R}_1={\bf I}_M$ is a positive definite matrix. 

Now, assume that ${\bf R}_n$ has all columns of the Hadamard matrix $(M$ by $M)$ as its eigenvectors and that ${\bf R}_n$ is positive definite for some $n \geq 1$. 
Since we assumed that ${\bf R}_n$ is symmetric positive definite matrix, all its eigenvalues are positive. Denote by ${\bf h}_1, {\bf h}_2,...,{\bf h}_M$ the $M$ columns of the Hadamard matrix ${\bf H}$.  By this encoding scheme, we set the vector ${\bf \alpha}_n={\bf h}_{(n\hspace{1mm} mod \hspace{1mm} M)+1}$. Assume that $\lambda_n$ is the eigenvalue of ${\bf R}_n$ associated with the $\alpha_n$ eigenvector. We have
\[
{\bf \alpha}_n^T {\bf R}_n {\bf \alpha}_n=||{\bf \alpha}_n||^2\lambda_n=M \lambda_n
\]
On the other hand, we also have
\[
\sum_{t=1}^{M} \alpha_n^{(t)}\rho_n^{(t,m)}={\bf \alpha}_n^T{\bf \rho}_n^{(m)}
\]
where ${\bf \rho}_n^{(m)}$ is the $m$th column of the matrix ${\bf R}_n$.
Note that ${\bf R}_n^T={\bf R}_n$, this means that
\[
{\bf R}_n^T {\bf \alpha}_n=\lambda_n {\bf \alpha}_n
\]
Hence,
\[
{\bf \alpha}_n^T\rho_n^{(m)}=\lambda_n\alpha_n^{(m)}, \hspace{3mm} \forall m \in \{1,2,...,M\}.
\]
Moreover, observe that
\[
\sum_{t=1}^M \sum_{l=1}^M \alpha_n^{(t)} \alpha_n^{(l)} \rho_n^{(t,l)}={\bf \alpha}_n^T {\bf R}_n {\bf \alpha}_n
\]
Combining these results, we obtain:
\begin{equation}
\sum_{t=1}^{M} \alpha_n^{(t)}\rho_n^{(t,m)}=\lambda_n\alpha_n^{(m)}
\end{equation}
and
\begin{equation}
\sum_{t=1}^M \sum_{l=1}^M \alpha_n^{(t)} \alpha_n^{(l)} \rho_n^{(t,l)}=M \lambda_n
\end{equation}
Substitute $(25)$ and $(26)$ into $(11), (12), (14), (15)$, we obtain:
\[
E[X_n^{(m)} Y_n] = \mbox{cov}\left(X_n^{(m)},Y_n\right) 
\]
\begin{equation}
= P\sum_{t=1}^M \alpha_n^{(t)}\rho_n^{(t,m)}=P \lambda_n \alpha_n^{(m)}
\end{equation}
\[
E[Y_n^2] = \mbox{var}(Y_n)=P \sum_{t=1}^M \sum_{l=1}^M \alpha_n^{(t)} \alpha_n^{(l)} \rho_n^{(t,l)} + 1
\]
\begin{equation}
=P {\bf \alpha}_n^T {\bf R}_n {\bf \alpha}_n + 1 = P M\lambda_n + 1
\end{equation}
\begin{equation}
A_n^{(m)}=\frac{P\lambda_n\alpha_n^{(m)}}{MP\lambda_n+1}
\end{equation} 
\begin{equation}
B_n^{(m)}=P\left[1-\frac{P\lambda_n^2}{MP\lambda_n+1}\right]
\end{equation}
Observe that from the proof of the Theorem II above, then
\[
X_{n+1}^{(m)}=\sqrt{P}\frac{X_n^{(m)}-A_n^{(m)}Y_n}{\sqrt{B_n^{(m)}}}, \hspace{2mm} \forall m \in \{1,2,...,M\}.
\]
Then, we have
\[
\rho_{n+1}^{(m,k)}=\frac{E[X_{n+1}^{(m)} X_{n+1}^{(k)}]}{P}=
\]
\[
=\frac{P\rho_n^{(m,k)}+ A_n^{(m)} A_n^{(k)} E[Y_n^2]}{\sqrt{B_n^{(m)}B_n^{(k)}}}
\]
\[
-\frac{A_n^{(m)} E[X_n^{(k)} Y_n]+A_n^{(k)} E[X_n^{(m)} Y_n] }{\sqrt{B_n^{(m)}B_n^{(k)}}}
\]
Moreover, from (13) we know that
\[
E[X_n^{(m)}Y_n]=A_n^{(m)} E[Y_n^2]
\]
Therefore, 
\[
\rho_{n+1}^{(m,k)}=\frac{P\rho_n^{(m,k)}-A_n^{(m)}A_n^{(k)}E[Y_n^2]}{\sqrt{B_n^{(m)}B_n^{(k)}}}
\]
Replacing the results in $(27), (28), (29)$, and $(30)$ to this expression, we obtain
\[
\rho^{(m,k)}_{n+1}=\frac{1+MP\lambda_n}{1+P\lambda_n(M-\lambda_n)}\rho_n^{(m,k)}
\]
\begin{equation}
-\frac{P\lambda_n^2}{1+P\lambda_n(M-\lambda_n)}\alpha_n^{(m)}\alpha_n^{(k)}
\end{equation}
for all $m,k \in \{1,2,...,M\}$.\\

Since we assumed that ${\bf R}_n$ is symmetric positive definite matrix, hence ${\bf R}_n={\bf R}_n^{T}$, hence $\rho_n^{(m,k)}=\rho_n^{(k,m)}, \hspace{2mm} \forall k,m \in \{1,2,...,M\}$. Therefore, from $(31)$, it is easy to see that $\rho_{n+1}^{(m,k)}=\rho_{n+1}^{(k,m)}, \forall k,m \in \{1,2,...,M\}$. In other words, ${\bf R}_{n+1}$ is also a symmetric matrix. \\
Moreover, from $(31)$, we also have
\[
{\bf R}_{n+1}=\frac{1+MP\lambda_n}{1+P\lambda_n(M-\lambda_n)}{\bf R}_n-\frac{P\lambda_n^2}{1+P\lambda_n(M-\lambda_n)}{\bf \alpha}_n {\bf \alpha}_n^T
\]
Denote $ {\bf H}_n=\left[\begin{array}{cccc}{\bf \alpha}_n&{\bf \alpha}_{n+1}&\cdots&{\bf \alpha}_{n+M-1}\end{array}\right] $. We see that the columns of ${\bf H}_n$ creates $M$ linearly independent eigenvectors of the matrix ${\bf R}_n$.  Moreover, we also have
\[
{\bf H}_{n+1}^T{\bf R}_{n+1}{\bf H}_{n+1}=\frac{1+MP\lambda_n}{1+P\lambda_n(M-\lambda_n)}{\bf H}_{n+1}^T{\bf R}_n{\bf H}_{n+1}
\]
\begin{equation}
-\frac{P\lambda_n^2}{1+P\lambda_n(M-\lambda_n)}{\bf H}_{n+1}^T{\bf \alpha}_n {\bf \alpha}_n^T{\bf H}_{n+1}
\end{equation}

Note that since all columns of ${\bf H}_n$ are eigenvectors of the matrix ${\bf R}_n$, so all the columns of the matrix ${\bf H}_{n+1}$ are also eigenvectors of the matrix ${\bf R}_n$, hence we has the following eigenvalue decomposition
\begin{equation}
{\bf \Lambda}_n={\bf H}_{n+1} {\bf R}_n {\bf H}_{n+1}^T
\end{equation} where ${\bf \Lambda}_n$ is a diagonal matrix. \\
Moreover, we have
\begin{equation}
{\bf H}_{n+1}^T{\bf \alpha}_n {\bf \alpha}_n^T{\bf H}_{n+1}=[{\bf \alpha}_n^T{\bf H}_{n+1}]^T {\bf \alpha}_n^T{\bf H}_{n+1}
\end{equation}
where
\[
{\bf \alpha}_n^T{\bf H}_{n+1}= {\bf \alpha}_n^T\left[\begin{array}{cccc}{\bf \alpha}_{n+1}& {\bf \alpha}_{n+2}&\cdots&{\bf \alpha}_{n+M}\end{array}\right]
\]
\begin{equation}
=\left[\begin{array}{cccc}0&0&\cdots&M\end{array}\right]
\end{equation}

From $(32), (33), (34), (35)$, the matrix ${\bf H}_{n+1}^T{\bf R}_{n+1}{\bf H}_{n+1}$ must be a diagonal one since the right side of $(32)$ is a diagonal matrix. Hence, all columns of the matrix ${\bf H}_{n+1}$ are eigenvectors of the matrix ${\bf R}_{n+1}$. 

Assuming $\left(\lambda_n^{(1)},\lambda_n^{(2)},...,\lambda_n^{(M)}\right)$ are $M$ eigenvalues corresponding to eigenvectors which are columns of the matrix ${\bf H}_n$.
By this notation, we see that $\lambda_n=\lambda_n^{(1)}, \hspace{2mm} \forall n \in \mathbb{N}$.\\
Combining $(32), (33), (34)$, and $(35)$ we obtain
\begin{equation}
\lambda_{n+1}^{(k)}=\left\{\begin{array}{l}\frac{(1+MP\lambda^{(1)}_n)\lambda_n^{(k+1)}}{(1+P\lambda^{(1)}_n(M-\lambda^{(1)}_n)},\hspace{8mm} k=1,2,...,M-1\\\frac{\lambda^{(1)}_n}{1+P\lambda^{(1)}_n(M-\lambda^{(1)}_n)}, \hspace{10mm} k=M
\end{array}\right.
\end{equation}
for all $k \in \{1,2,...,M\}$.  

Since we assumed that all eigenvalues of ${\bf R}_n$ are positive (${\bf R}_n$ symmetric positive definite),  from $(36)$ we see that all eigenvalues of ${\bf R}_{n+1}$ are also positive. Note that we also confirmed that ${\bf R}_{n+1}$ is symmetric above, therefore ${\bf R}_{n+1}$ is a symmetric positive definite matrix. In short, if ${\bf R}_n$ is a symmetric positive definite matrix and has all columns of the Hadamard matrix as its eigenvectors, then ${\bf R}_{n+1}$ has all these properties. This concludes our proof by induction.

According to the Lemma 1 [12] (or see the Appendix below), the sequence $\lambda_n^{(1)}$ converges (or can be forced to converge) to a fixed point $\lambda^*$,  which is the biggest positive solution in $[1,M]$ of the following equation
\begin{equation}
(PM\lambda+1)^{M-1}=[P\lambda(M-\lambda)+1]^M
\end{equation}
From (27), (28) we obtain 
\[
 \liminf_{n \rightarrow \infty} \frac{\left(\sum_{t=1}^M \alpha_n^{(t)}\rho_n^{(t,m)}\sqrt{P_t}\right)^2}{\sum_{t=1}^M\sum_{l=1}^M \alpha_n^{(t)}\alpha_n^{(l)}\rho_n^{(t,l)}\sqrt{P_tP_l}+1}
 \]
 \[
 =\liminf_{n\rightarrow \infty}\frac{P\lambda_n^2}{PM\lambda_n+1}=\frac{P(\lambda^*)^2}{PM\lambda^*+1}
\]
Since $1 \leq \lambda^* \leq M$, we have
\[
 \displaystyle{0< \frac{P(\lambda^*)^2}{PM\lambda^*+1}<1}
 \]
Therefore, the constraints in the Theorem II is satisfied. Applying the result of this theorem, we have any rate less than
\[
R_m^*=-\frac{1}{2}\limsup_{n\rightarrow \infty}\log\left[1-\frac{P\lambda_n^2}{MP\lambda_n+1}\right]
\]
\[
=-\frac{1}{2}\log\left[1-\frac{P(\lambda^*)^2}{MP\lambda^*+1}\right]
\] 
is achievable,  for all $m=1,2,...,M$. Hence, any sum rate which is less than
\[
\sum_{m=1}^M R_m^*=\frac{M}{2}\log\left(\frac{1+MP\lambda^*}{1+P\lambda^*(M-\lambda^*)}\right)
\]
\[
=\frac{1}{2}\log\left(\left[\frac{1+MP\lambda^*}{1+P\lambda^*(M-\lambda^*)}\right]^M\right)=\frac{1}{2}\log\left(1+PM\lambda^*\right)
\]
is achievable, where $\lambda^*$ is solution in the $[1,M]$ of the equation $(37)$. This result coincides with the formula (68) in [14]. The paper [16] proves that this achievable sum rate is optimal for the class of linear feedback coding. \\

\section{Conclusion}
A posterior matching based encoding-decoding strategy for general Gaussian MAC with feedback was proposed, and achievable rate region, error performance were drawn.  Finally, we analyzed error performance of the proposed posterior encoding scheme and showed that the \emph{time-varying posterior matching scheme} and variable rate decoding ideas can be applied to Gaussian MAC and obtain optimal performances. Specifically, the proposed encoding scheme achieves the capacity of two-user feedback Gaussian MAC as well as linear-feedback sum-rate for symmetric Gaussian MAC with feedback where the number of users is arbitrary. Moreover, by the encoding scheme's structure, which uses the spreading codes like the Hadamard matrix, our encoding scheme can be directly applied to CDMA systems with feedback. Finally, by analyzing all arguments in the theorem I, the \emph{time-varying posterior matching scheme} approach in this paper might be applied for other Gaussian and non-Gaussian multiuser channels to achieve optimal performances. 
\appendix [Proof of Lemma II]
\begin{IEEEproof}
We use the same line argument as Lemma 1 [4]. Assume we are given a transmission scheme with $M$ transmission functions $g_n^{(m)}$ and a decoding rule which are known to achieve the rate vector $(R^*_1,R^*_2,...,R^*_M)$.  For simplicity, we assume that the decoding rule is fixed rate $(i.e., |\Delta^{(m)}(y^n)|=2^{-nR^*_m}$ for all $y^n$), since any variable rate decoding rule can be easily mapped into a fixed rate rule that achieves the same rate vector.  It is easy to see that in order to prove that the above translates into achievability for some rate vector $\{(R_1,R_2,...,R_M): R_1< R_1^*, R_2< R_2^*,...,R_M < R_M^*)\}$ in the standard framework, it is enough to show we can find $M$ sequences $\Gamma^{(m)}_n=\left\{\theta_{i,n}^{(m)} \in (0,1)\right\}_{i=1}^{\lfloor{2^{nR_m}} \rfloor},$ and such that we have the uniform achievability over $\Gamma_n^{(m)}$, i.e.,
\[
\lim_{n\rightarrow \infty} \max_{\theta_m \in \Gamma_n^{(m)}} \mathbb{P}(\theta_m \notin \Delta_n^{(m)}(Y^n)|\Theta_m=\theta_m)=0 
\]
We now show how $\Gamma_n^{(m)}$ can be constructed for any $R_m < R_m^*$. Let $p_n^{(m)}(e)$ be the (average) error probability associated with our scheme and the fixed rate vector $(R_1^*, R_2^*,...,R_M^*)$. Define 
\[
A_n^{(m)}=\left\{\theta_m \in (0,1): \mathbb{P}(\Theta_m \notin \Delta_n^{(m)}(Y^n)|\Theta_m=\theta_m) \right. \\
\]
\[
\left.> \sqrt{p_n^{(m)}(e)}\right\}
\]
and write 
\[
p_n^{(m)}(e)=\int \mathbb{P}(\Theta_m \notin \Delta_n^{(m)}(Y^n)|\Theta_m=\theta_m)d\theta_m 
\]
\[
> \sqrt{p_n^{(m)}(e)}\int {\bf 1}_{A_n^{(m)}}(\theta_m)d\theta_m
\]
and so we have that $\int {\bf 1}_{A_n^{(m)}}(\theta_m)d\theta_m < \sqrt{p_n^{(m)}(e)}$. It is now easy to see that if we want to select $\Gamma_n^{(m)}$ such that $\Gamma_n^{(m)}\cap A_n^{(m)}=\emptyset$, and also $\theta_{i+1,n}^{(m)}-\theta_{i,n}^{(m)} \geq 2^{-nR_m^*}$, then a sufficient condition is that  $\frac{1}{|\Gamma_n^{(m)}|}\left(1-\sqrt{p_n^{(m)}(e)}-\tau_n^{(m)}\right) \geq 2^{-nR_m^*} $ for some positive $\tau_n^{(m)}\rightarrow 0$. This condition can be written as
\[
\frac{1}{n}\log|\Gamma^{(m)}_n| \leq R_m^* +\frac{1}{n}\log\left(1-\sqrt{p_n^{(m)}(e)}-\tau_n^{(m)}\right)
\]
\[
=R_m^*+o(1)
\]
At the same time, we also have by definition
\[
\lim_{n\rightarrow \infty} \max_{\theta_m \in \Gamma_n^{(m)}} \mathbb{P}(\theta_m \notin \Delta_n^{(m)}(Y^n)|\Theta_m=\theta_m)
\]
\[
\leq  \lim_{n\rightarrow \infty} \sqrt{p_n^{(m)}(e)}=0, \hspace{2mm} \forall m \in \{1,2,...,M\}
\]
\end{IEEEproof}
\appendix [Forcing the sequence convergence]
\begin{IEEEproof}
We use the similar arguments as the Appendix A [14]. It is difficult to prove that the recursion (36) is convergence. However, we know that for $n \geq M+1$
\[
\lambda_{n+1}^{(1)}=\left[\frac{1+PM\lambda_n^{(1)}}{1+P\lambda_n^{(1)}(M-\lambda_n^{(1)})}\right] \lambda_n^{(2)} = 
\]
\[
=\left[\frac{1+PM\lambda_n^{(1)}}{1+P\lambda_n^{(1)}(M-\lambda_n^{(1)})}\right] \left[\frac{1+PM\lambda_{n-1}^{(1)}}{1+P\lambda_n^{(1)}(M-\lambda_{n-1}^{(1)})}\right]\lambda_{n-1}^{(2)}
\]
\[
= \left[\prod_{k=0}^{M-1}\left( \frac{1+PM\lambda_{n-k}^{(1)}}{1+P\lambda_{n-k}^{(1)}(M-\lambda_{n-k}^{(1)})}\right)\right]
\]
\[
\times \left[\frac{\lambda_{n-M}^{(1)}}{1+P\lambda_{n-M}^{(1)}(M-\lambda_{n-M}^{(1)}}\right]
\]
Therefore, if the sequence $\lambda_n^{(1)}$ is convergent, it will converges to the solution of the following equation:
\[
(P\lambda+1)^{M-1}=[((P/M)\lambda(M-\lambda)+1]^M
\]
It is easy to show that this equation has at least one solution in the $[1, M]$. Set $\lambda_1 =\lambda^*$ to be the biggest solution of this equation. We also set
\[
\lambda_{m+1} =\frac{1+(P/M)\lambda_1(M-\lambda_1))}{1+P\lambda_1}\lambda_m, \hspace{1mm} m=1,2,\cdots, M-1
\]
We will force the recursion (36) to yield  the desired values $\lambda_M^{(1)}=\lambda_1, \lambda_M^{(2)}=\lambda_2,\cdots, \lambda_M^{(M)}=\lambda_M$ at time $n=M$.

To perform this forcing, we replace $P$ by $P_n$ at time $n$ and apply (36) $M-1$ times to get
\[
\lambda_M^{(m)} =\frac{1}{D}\prod_{n=1, n\neq m-1}^{M-1} [P_n\lambda_n +1]
\]
for all $m$, where
\[
D=\prod_{n=1}^{M-1}[(P_n/M)\lambda_{1n}(M-\lambda_n^{(1)}+1]
\]
We thus have, for $n=1,2,\cdots, M-1$
\[
\lambda_{M}^{(1)}/{\lambda_M^{(n+1)} }= P_n \lambda_n^{(1)} +1
\]
Solving for $P_n$, we have
\[
P_n = \frac{1}{\lambda_{n}^{(1)}}\left(\lambda_M^{(1)}/{\lambda_M^{(n+1)}}-1\right)
\]
where $\lambda_n^{(1)}$ can be computed recursively via $\lambda_1^{(1)}= 1$ and
\[
\lambda_n^{(1)} = \prod_{k=1}^{n-1}\frac{P_k\lambda_k^{(1)} +1}{(P_k/M) \lambda_k^{(1)}(M-\lambda_m^{(1)})+1} 
\]
\end{IEEEproof}
\section*{Acknowledgment}
The author would like to thank David J. Love, Purdue University, West Lafayette, United States for introducing me to the communications with feedback. The author also would like to thank anonymous reviewers for their useful suggestions to improve the manuscript.  

\end{document}